\documentclass[twocolumn,showpacs,amssymb,aps, prd]{revtex4}
\usepackage[T1]{fontenc}
\usepackage{amsmath}
\usepackage{amssymb}
\usepackage{mathrsfs}
\usepackage{graphicx}
\usepackage{amsmath}
\usepackage{graphicx}
\usepackage{amssymb}
\usepackage{mathrsfs}
\newtheorem{thm}{Theorem}
\newtheorem{prop}{Proposition}

\newcommand{\ud}{\mathrm{d}}
\def\mby{\mathfrak{m}_{\rm{BY}}}

\newcommand{\be}{\begin{equation}}
\newcommand{\ee}{\end{equation}}
\newcommand{\bee}{\begin{equation*}}
\newcommand{\eee}{\end{equation*}}

\newcommand{\beq}{\begin{eqnarray}}  \newcommand{\eeq}{\end{eqnarray}}

\begin{document}

\title{Brown-York mass and the hoop conjecture in non-spherical massive systems}

\author{Edward Malec}\email{malec@th.if.uj.edu.pl}
\affiliation{
Institute of Physics, Jagiellonian University, \L ojasiewicza 11,  30-438 Krak\'ow, Poland
}

\author{Naqing Xie}\email{nqxie@fudan.edu.cn}
  \affiliation{School of Mathematical Sciences, Fudan University, Shanghai, China}

\date{\today}

\begin{abstract}
We discuss the relation between the concentration of the Brown-York mass and the formation of trapped surfaces in non-spherical massive systems. In particular, we formulate and prove a precise version of the Thorne hoop conjecture in conformally flat three-geometries sliced by equipotential   foliation leaves. An intriguing relationship between the total rest mass and the Brown-York mass is shown. This is a further investigation of the previous work on the Brown-York mass hoop conjecture in spherical symmetry.
\end{abstract}

\maketitle

\section{Introduction}
The Brown-York quasi-local mass \cite{BY} is one of many mass concepts developed in last decades.   Its potential has  been revealed  in \cite{OTXM}, where   the Brown-York mass has been used in order to prove the trapped surface conjecture \cite{Sei} in spherically symmetric geometries. The trapped surface conjecture states  that  large mass enclosed in a small volume  has to be trapped, and it constitutes an attempt  to  concretize a loose idea expressed by Thorne in his hoop conjecture \cite{Thorne}.

In this paper we demonstrate that the Brown-York mass is useful in proving the hoop conjecture in certain classes of non-spherical geometries.   They include systems having an equipotential surface foliation, that is convex in a certain sense.   We present   sufficient conditions for the existence of trapped surfaces.  It appears convenient to split the consideration into two stages. In the first step one deals with 2-surfaces that satisfy an integral condition - that they are  averaged trapped surfaces \cite{HW}. In the second step one finds additional conditions that ensure the pointwise trapping of averaged trapped surfaces.

The paper is organized as follows. Section \ref{S1} gives a concise historical account on  the hoop conjecture. Section \ref{S2} contains the description of the formalism and needed definitions.  We propose various sufficient and necessary conditions for the existence of averaged trapped surfaces in general settings.  This is done  in terms of the reference geometry. Section {\ref{S4}   defines an equipotential  foliation, assuming conformally flat geometry. We provide therein necessary and  sufficient conditions for averaged trapped surfaces in terms of the original physical geometry.  As an aside, but important result,  we show that the Brown-York mass is not larger than the total rest mass. The question whether an averaged trapped surface is indeed pointwise trapped is examined in Section \ref{S3}.  There is an  extra mass term required to balance the non-sphericality, that can be written down explicitly. The last Section summarizes obtained results.

\section{Hoop conjecture}\label{S1}
There is a folk belief in general relativity that if matter is sufficiently concentrated into a finite volume, the gravitational system ultimately has to collapse to a black hole. Thorne   proposed a hoop conjecture (HC) \cite{Thorne} which states:
\begin{quote} Horizons form when and only when a mass $M$ gets compacted into a region whose circumference in EVERY direction satisfies $C \lesssim 4 \pi M$. \end{quote}
His conjecture deals with global event horizons and  the `circumference' and the mass are deliberately
left unspecified.  Notice, however, that in the Schwarzschild spacetime we have the equality:  $C= 4 \pi M$. In this case $M$ is the asymptotic mass \cite{ADM}. Seifert \cite{Sei}   formulated  the more concrete trapped surface conjecture (TSC), according to which    massive singularities have to be  surrounded by a closed trapped
2-surface.  This is an easier concept, because trapped surfaces are local in time. Proving the HC would require the study of the full history of a spacetime, while in order to prove the TSC one  needs only to consider a single Cauchy slice.

There have been many attempts to prove   the HC/TSC.  In the early period the concentration of matter was assumed in spherically symmetric spacetimes  \cite{BMO88,BMO89,Za92,Za93,M94,Ea,JO1,JO2,MO04}. Early results have been reviewed in \cite{M91a}.  Recently Khuri   \cite{Kh}   applied in this context  the generalized Jang equation \cite{BK}.   Schoen and Yau dealt    with nonsymmetric spacetimes \cite{SY2}. Their sufficient condition for the formation of trapped surfaces   required a special spacelike foliation of a spacetime  with large extrinsic curvature. Within a single Cauchy slice, assuming the matter density to be large on a 'large region',  trapped surfaces have to form. That   was   a consequence of the blow-up analysis of the Jang equation \cite{Ja} for an asymptotically flat initial data set \cite{SY}.  In later studies the TSC has been proved in special classes of systems, with matter \cite{KM, M91, M91a} or in vacuum \cite{BOM, Ea}. Some of the recent development has been reviewed in \cite{Sz}.

In spherical symmetric systems one can measure their 'size'  by the circumference. It is reasonable to take $C=2\pi R$ where $R$ is the Schwarzschild or the areal radius of the surface in question, i.e. $R=\sqrt{\mbox{Area}/4\pi}$.  Then one can prove a precise statement of the hoop conjecture  using the Brown-York mass as the mass measure   \cite{OTXM}. The theorem says that if $C<2\pi \mby$, then the surface is trapped.  There exists a discrepancy between the   $4\pi$ in Thorne's HC and the coefficient $2\pi$ in \cite{OTXM}.  This can be traced back to the fact that at the   horizon  of the Schwarzschild spacetime the Brown-York mass is equal to $R$, the area radius of the horizon. That is $\mby = 2M$, where $M$ is the asymptotic mass.

\section{Averaged Trapped Surfaces}\label{S2}
Let $(\Omega^3,g,K)$ be a subset of a Cauchy slice for the Einstein field equations. Here $g$ is the 3-metric of a Cauchy hypersurface and $K$ stands for its extrinsic curvature. We assume that $\Omega$ is time-symmetric, i.e. it lies in a  totally geodesic Cauchy hypersurface, $K\equiv 0$.

In this paper, we concentrate on the case of $\Omega$ being a compact domain with boundary and since it is time symmetric, we use the Brown-York mass as our measure of the mass within $\Omega $.

Assume further that the boundary $\Sigma=\partial \Omega$ is a topological 2-sphere. There exists a unit normal $n$ (directed outward) to $\Sigma $; its divergence $\nabla_i   n^i$ is equal to the mean curvature $k$.  Here $\nabla_i  $ denotes the covariant derivative with respect to the 3-metric $g$.  The sign of $k$ has an important physical meaning. Take a bundle of outgoing null rays, normal to $\Sigma $. If $k>0$ along $\Sigma $, then the bundle is divergent; while if $k<0$, then the null rays must converge. If $k<0$ everywhere along
 $\Sigma $, then the two-surface is said to be {\it trapped}. Trapped surfaces do not exist  in the Euclidean
geometry and their presence is associated with strongly curved geometries.

On the other hand, the first derivative of the area of $\Sigma$ with respect to the uniform normal deformation gives the total mean curvature,
\be
H(\Sigma)=\int_\Sigma k\ud\Sigma.\ee
The concept of a trapped surface is purely local,  but   it  appears   useful
to deal with surfaces that are trapped in the average:

\noindent {\bf Definition}
A surface $\Sigma$ is called an {\it averaged trapped surface} (ATS) if $H(\Sigma)$ is negative.

Assume further that $\Sigma$ has positive Gauss curvature and thus can be isometrically embedded into the Euclidean space $\mathbb{R}^3$, i.e. $i: \Sigma \hookrightarrow i(\Sigma) \subset \mathbb{R}^3$.  This isometric embedding is called the Weyl embedding and it is unique up to a rigid motion in $\mathbb{R}^3$ \cite{Ni}.

Then the Brown-York mass \cite{BY} is defined as
\be
\mby(\Sigma,g)=\frac{1}{8\pi}\int_\Sigma (k_0-k)\ud\Sigma\ee
where $k$ is the mean curvature of $\Sigma$ with respect to the physical metric $g$ and $k_0$ is that of $i(\Sigma)$ with respect to the Euclidean metric. Note that $k_0$ is completely determined by the intrinsic 2-metric on the surface $\Sigma$ but does not depend on the extrinsic geometry how $\Sigma$ bends in $\Omega$.

The above definition implies in a straightforward way the {\it important} proposition.

\begin{prop}\label{prop1}The surface $\Sigma$ is an ATS if and only if
\be
\mby(\Sigma,g) > \frac{1}{8\pi}\int_\Sigma k_0 \ud\Sigma.
\ee
\end{prop}
It is interesting that here the integral  and the mean curvature $k_0$  in  Proposition 1  are in the Euclidean space. One can employ well known geometric estimates and re-express the proposition
in a number of ways. This is done in the remainder of this section.
The total mean curvature $\int_\Sigma k_0 \ud \Sigma $ represents an 'averaged size' of a solid convex body in the Euclidean space \cite{PS}. Suppose that a compact oriented convex surface $\Sigma$ lies in $\mathbb{R}^3$. Let $x_0$ be a fixed point enclosed by $\Sigma$. The Minkowski integral formula \cite[Lemma 6.2.9, Page 136]{Kl} gives
\be\label{MF}
\int_\Sigma \frac{k_0}{2} \ud \Sigma = \int_{\Sigma}K(x)<n(x),X(x)-x_0>_{\mathbb{R}^3}\ud\Sigma.\ee
Here $K(x)$ is the Gauss curvature and $X(x)$ is the position vector of $\Sigma$ in $\mathbb{R}^3$, $n(x)$ is the unit normal at $X(x)$ and $<\cdot,\cdot>$ denotes the Euclidean inner product.

Recall that $\Sigma$ is a topological sphere. By the Gauss-Bonnet theorem $\int_\Sigma K(x)\ud \Sigma=2\pi\chi(S^2)=4\pi$, it gives an upper bound of the right hand side of (\ref{MF}), $4\pi\sup_{x\in \Sigma}|X(x)-x_0|$.

If we measure the 'size' of a surface by looking at the position vector of its image when embedded isometrically into $\mathbb{R}^3$, then we have
\begin{thm}\label{T1} ({\it Sufficient Condition for an ATS}.) If \be
\mby(\Sigma,g) > \sup_{x\in \Sigma}|X(x)-x_0|,\ee then $\Sigma$ is an ATS.
\end{thm}

Another upper bound of the total mean curvature is given by the Blaschke cap body inequality, cf. Page 387 in \cite{Schn}. Let $V$ be a compact convex body in $\mathbb{R}^3$. Then
\bee
\mbox{Area}(\partial V)\geq \sqrt{3 \mbox{Vol}(V)\int_\Sigma \frac{k_0}{2} \ud(\partial V)}\eee
where $k_0$ is the mean curvature of the boundary $\partial V$.
This leads to the following theorem.
\begin{thm}\label{T2}({\it Sufficient Condition for an ATS}.) If \be
\mby(\Sigma,g) > \frac{1}{4\pi} \frac{\big(\mbox{Area}(\Sigma)\big)^2}{3 \mbox{Vol}(\Omega_0)},\ee
then $\Sigma$ is an ATS. Here $\Omega_0$ is the body in $\mathbb{R}^3$ enclosed by the image of $\Sigma$ via the (unique) Weyl embedding.\end{thm}

At this stage we have a hybrid picture. The Brown-York mass lives in a physical space while the upper bounds are given in the reference space. Again things are easy in the spherically symmetric case \cite{OTXM}, when  $k_0=2/R$, where $R$ is the areal radius.  In general Riemannian geometries, life becomes harder. It is difficult to define a workable concept of a  'circumference' \cite{Fl,Se}. Fortunately, one  finds a quantitative link between the total mean curvature in the reference space $\int_\Sigma k_0\ud\Sigma$ and the original physical geometric data in a class of foliations of conformally flat 3-manifolds. The details will be discussed in the next section.

To provide necessary conditions for an ATS, we need the lower bound estimates of the total mean curvature $\int_\Sigma k_0 \ud\Sigma$. There are two candidates both of which are in terms of intrinsic 2-geometry of the surface. One is given by the classical geometric inequality \cite{Schn} and the other one is given by the Birkhoff invariant of the intrinsic 2-metric \cite{JCAP}.

\begin{thm}\label{T3} ({\it Necessary Condition for an ATS}.) Assume that $\Sigma$ is an ATS, then\\
\be
\mbox{(1)}\ \mby(\Sigma,g) > \sqrt{\frac{\mbox{Area}(\Sigma)}{4\pi}};\ee
\be
\mbox{(2)}\ \mby(\Sigma,g) > \frac{1}{8\pi}\cdot4\beta=\frac{\beta}{2\pi}.\ee
Here $\beta$ is the Birkhoff invariant of the surface $\Sigma$.\end{thm}

\section{Conformally Flat Geometries}\label{S4}
Herein we shall  investigate the following concrete class of three-spaces. Assume that\\
\noindent (1) $g$ is conformally flat, $g_{ab}=f^4\hat{g}_{ab}$ where $\hat{g}_{ab}$ is the standard Euclidean metric.\\
\noindent (2) There is an equipotential foliation on $\Omega$,
\be\label{foli}
g=f^4(\sigma)[\hat{g}_{\sigma\sigma}\ud\sigma^2+\hat{g}_{ij}\ud x^i \ud x^j]\ (i,j=2,3)\ee
where $\sigma\geq 0$ and $\sigma$ foliates the level surfaces of $f$ which are assumed to be convex, and $x^2$ and $x^3$ are quasi-angle variables.\\
\noindent (3) $\Sigma=\{\sigma=\sigma_0\}$.

Thus, $\hat{n}=(\hat{n}_\sigma,0,0)$ and $\hat{n}_\sigma=\sqrt{\hat{g}_{\sigma\sigma}}$. The conformal factor $f$ satisfies the elliptic equation $\hat{\Delta}f=-2\pi\rho f^5$, which is  the Hamiltonian constraint for momentarily static initial data of  the Einstein equations. The energy density $\rho $ is nonnegative due to  energy conditions.

\noindent {\it Remark:}
In order to detect whether a surface $\Sigma$ is trapped or not, one only needs the geometry in a neighborhood of $\Sigma$ in $\Omega$, or within $\Sigma $.
The dominant view nowadays is that trapped surfaces are of physical interest  because of their  roles in  the proof of scenarios of the cosmic censorship. That demands that   $\Omega$ constitutes a domain of an asymptotically flat Cauchy slice and hence one should  assume that  the equipotential surface foliation, that covers $\Omega $,  is extendible onto the entire slice with the asymptotic condition $f(\infty)=1$. That in turn implies  that $f|_{\Sigma}\geq 1$ by the maximum principle.

We emphasize that $\Sigma$ refers to the $\sigma=\sigma_0$ surface with induced metric $f^4(\sigma_0)(\hat{g}_{ij}\ud x^i \ud x^j)$. Denote by $\hat{\Sigma}$ the $\sigma$-constant surface with induced metric $\hat{g}_{ij}\ud x^i \ud x^j$. Let $k$ be the mean curvature of $\Sigma$ with respect to the physical metric $g$ and let $\hat{k}$ be the mean curvature of $\hat{\Sigma}$ with respect to the Euclidean metric $\hat{g}_{\sigma\sigma}\ud\sigma^2+\hat{g}_{ij}\ud x^i \ud x^j$.

We isometrically embed $\Sigma$ into the reference space $f^4(\sigma_0)[\hat{g}_{\sigma\sigma}\ud\sigma^2+\hat{g}_{ij}\ud x^i \ud x^j]$ which is also Euclidean. Then it gives a relation between $k_0$ and $\hat{k}$, i.e. $k_0=\hat k/ f^2(\sigma_0)$ and that of the induced area forms is
$\ud\Sigma=f^4(\sigma_0)\ud\hat{\Sigma}$.

Below we shall write down some criteria for ATS's, obtained in Section \ref{S2},  in terms of the geometry of the physical space $(\Omega,g)$.

(i) Proposition \ref{prop1} states that $\Sigma $ must be an ATS if   $\mby(\Sigma,g)> \int_\Sigma k_0 \ud \Sigma $; but   $  \int_\Sigma k_0 \ud \Sigma =f^2(\sigma_0)\int_\Sigma \hat  k \ud \hat \Sigma$  represents an 'averaged areal size' $R_{Av}$ of a body enclosed by the (convex) 2-surface  $\Sigma $. The sufficiency condition states simply  $\mby(\Sigma,g) >R_{Av}(\Sigma )$.

(ii) In the same way one may also rewrite Theorem \ref{T1} as:
If
$\mby(\Sigma,g) >R_{sup}(\Sigma )$, where $R_{sup}(\Sigma ) := f^2(\sigma_0) \sup_{x\in \hat\Sigma}|\hat{X}(x)-x_0|,$ then $\Sigma$ is an ATS. Here $\hat{X}$ is the position vector of the surface $\hat{\Sigma}$ in the Euclidean space $\hat{g}_{\sigma\sigma}\ud\sigma^2+\hat{g}_{ij}\ud x^i \ud x^j$.

As a consequence of the uniqueness of the Weyl embedding, the lower bounds given in Theorems \ref{T2} and \ref{T3} are completely determined by the intrinsic 2-geometry on the surface. They are the same no matter calculated either in the physical space $(\Omega,g)$ or in the reference Euclidean space. In particular,

(iii) If we define the areal radius of $\Sigma$ as $R_S:=\sqrt{\mbox{Area}(\Sigma)/4\pi}$, then the first condition in Theorem \ref{T3} becomes $\mby(\Sigma,g) > R_S$.

(iv) In the second condition of Theorem \ref{T3}, the Birkhoff invariant $\beta$ is the minimum length of a closed string being slipped over the 2-surface \cite{Birk}. One defines the Birkhoff radius $R_B:=\beta/2\pi$ and then the condition becomes $\mby(\Sigma,g)>R_B$.

We have introduced the above four size measures for $\Sigma$. Notice that in spherical geometries and for a round sphere $\Sigma$ centered at the symmetry center, all these measures coincide, $R_{Av}=R_{sup}=R_S=R_B$.

The physical and 'embedded'  mean  curvatures  along $\Sigma$  are related as
\be
k|_\Sigma=\frac{\hat{k}|_{\hat\Sigma(\sigma=\sigma_0)}}{f^2(\sigma_0)}+\frac{4}{f^3(\sigma_0)}\hat{n}^a\hat{\nabla}_a f|_{\hat\Sigma(\sigma=\sigma_0)}.\ee

There is a simple calculation that allows us to give an upper bound onto the Brown-York mass $\mby(\Sigma,g)$ by the total rest mass $M(\Sigma)=\int_\Omega\rho \ud vol_g$ within $\Sigma $.  Indeed,
\be\label{BYrest}\begin{aligned}
\mby(\Sigma,g)&=\frac{1}{8\pi}\int_\Sigma\frac{\hat{k}}{f^2(\sigma_0)}f^4(\sigma_0)\ud\hat{\Sigma}\\
&\ \ -\frac{1}{8\pi}\int_\Sigma\big(\frac{\hat{k}}{f^2(\sigma_0)}+\frac{4}{f^3(\sigma_0)}\hat{n}^a\hat{\nabla}_a f \big)f^4(\sigma_0)\ud\hat{\Sigma}\\
&=-\frac{4f(\sigma_0)}{8\pi} \int_\Sigma \hat{n}^a\hat{\nabla}_a f \ud\hat\Sigma =-\frac{f(\sigma_0)}{2\pi } \int_\Omega\hat{\Delta}f\ud vol_{\hat{g}}\\
&=\int_\Omega f(\sigma_0)\rho f^{5} f^{-6}\ud vol_g=\int_\Omega\rho \frac{f(\sigma_0)}{f}\ud vol_g\end{aligned},\ee
where $\hat{\Delta}f=-2\pi\rho f^5$.
If we assume the dominant (or weak)  energy condition, i.e. $\rho\geq 0$, then $f$ is a superharmonic function (with respect to the Euclidean metric) and the maximum principle yields that for any $x \in \Omega$, $f(x)\geq f|_{\Sigma}=f(\sigma_0)$. Therefore,
\begin{thm}\label{restBY}
One has $\mby(\Sigma,g)\le M(\Sigma)$ where $M(\Sigma)=\int_\Omega\rho \ud vol_g$.
\end{thm}
\noindent {\it Remark: } The total rest mass has been employed in \cite{BMO88,BMO89,BM89,M91} in the derivation of sufficient conditions for ATS's and further TS's under certain additional conditions. The conditions therein are of the form $M(\Sigma) > D(\Sigma)$ where $D(\Sigma)$ is a certain 'size measure' coming from an upper bound of the geometric size of the domain enclosed by $\Sigma$. As a corollary of Theorem \ref{restBY}, if $\mby(\Sigma,g)>  D(\Sigma )$, then $\Sigma $ must be an ATS or TS. One is expecting to find a refined size measure $D^\prime(\Sigma)$ (which is smaller than $D(\Sigma)$) for the Brown-York mass. We shall do it in the next section.

If $\Sigma$ is a marginally trapped massive shell where the derivative of $f$ has a discontinuity, then Eq.(\ref{BYrest}), as a consequence of integration by parts, is no longer valid. But in spherical symmetry, one shows that the total rest mass equals twice of the asymptotic mass \cite[Eq.(1.14)]{BMOM1990}. This value also agrees with the Brown-York mass.

\section{From Averaged Trapped Surface to Trapped Surface}\label{S3}
In spherically symmetric geometries, if we take a spherical two-surface $\Sigma $ centered at the symmetry center, then its  mean curvature becomes a constant. That means that  it is trapped if and only if it is an ATS. This is the situation considered in \cite{OTXM}. In non-spherical geometries, the existence of an ATS is {\it not} sufficient to make use of the Penrose singularity theorem \cite{Pe} to conclude that a black hole has to develop. In this section, we formulate certain additional conditions which guarantee that an 'ellipsoidal' ATS is indeed trapped. We would ask how much Brown-York mass compacted into the system can produce a pointwise TS. The reasoning is analogous to that used in \cite{BM89,M91,M91a,KM}.

Suppose that $\Sigma$ is not a TS, there must be at least a point on which the mean curvature $k$ is nonnegative. Then the maximal value of $n_\sigma k$ must be nonnegative and hence
\be
\int_\Sigma n^\sigma(n_\sigma k)_{max}\ud\Sigma\geq0.\ee
There is no summation for $\sigma$ here. Instead, $n^\sigma$ denotes the particular $\sigma$-component of the unit normal in the equipotential foliation (\ref{foli}). Then one must have
\be
\frac{1}{8\pi}\int_\Sigma n^\sigma[(n_\sigma k)_{max}-n_\sigma k]d\Sigma+\frac{1}{8\pi}\int_\Sigma k_0 \ud\Sigma \geq \mby(\Sigma,g).\ee
Equivalently, we have
\begin{prop}If
\be\label{TSS}
\mby(\Sigma,g) > \frac{1}{8\pi}\int_\Sigma n^\sigma[(n_\sigma k)_{max}-n_\sigma k]\ud\Sigma+\frac{1}{8\pi}\int_\Sigma k_0 \ud\Sigma,\ee
then $\Sigma$ must be a pointwise TS.\end{prop}

Now we apply Eq.(12) in \cite{M91}:
\be
\partial_\sigma (\int_{\hat{\Sigma}}\hat{k}\ud\hat{\Sigma})=2\int_{\hat{\Sigma}}\hat{K}\hat{n}_\sigma \ud\hat{\Sigma}:=8\pi C(\sigma).\ee

Integrating from $0$ to $\sigma_0$, we have
\be
\begin{aligned}
\frac{1}{8\pi}\int_\Sigma k_0 d\Sigma&=\frac{1}{8\pi }\int_{\hat{\Sigma}(\sigma=\sigma_0)}\frac{\hat{k}}{f^{2}(\sigma_0)}f^{4}(\sigma_0)\ud\hat{\Sigma}\\
&=\frac{f^{2}(\sigma_0)}{8\pi}\int_{\hat{\Sigma}(\sigma=0)}\hat{k}\ud\hat{\Sigma}+f^{2}(\sigma_0)\int_0^{\sigma_0} C(s)\ud s.\end{aligned}\ee
Note that the $\{\sigma=0\}$ 'surface' is the set of points for which the conformal factor $f$ achieves its maximal value.

One may find an upper bound of the first term, Eq.(18) in \cite{Malec22},
\be
\frac{f^2(\sigma_0)}{8\pi}\int_{\hat{\Sigma}(\sigma=0)}\hat{k}\ud\hat{\Sigma}
\leq \frac{f^2(\sigma_0)\pi }{4}\sup l(S(0)). \ee
Here $\sup l(S(0))$ is the largest flat radius of the disk on which the conformal factor $f$ achieves its maximal
value.

Finally, we have arrived at
\begin{thm} ({\it Sufficient condition for a pointwise TS.}) If \be\label{suff}\begin{aligned}
\mby(\Sigma,g) >  &\ \frac{1}{8\pi}\int_\Sigma n^\sigma[(n_\sigma k)_{max}-n_\sigma k]\ud\Sigma\\
&+\frac{f^2(\sigma_0)\pi }{4}\sup l(S(0))\\
&+f^{2}(\sigma_0)\int_0^{\sigma_0} C(s)\ud s,\end{aligned}\ee then $\Sigma$ is a pointwise TS. \end{thm}
The physical significance of this theorem is as follows.\\
\noindent (1) The line integral term $f^{2}(\sigma_0)\int_0^{\sigma_0} C(s)\ud s$ represents an appropriate 'size' of the hoop for mass concentration.\\
\noindent (2) The surface integral term $\int_\Sigma n^\sigma[(n_\sigma k)_{max}-n_\sigma k]\ud\Sigma/8\pi$ reflects the 'boundary effect' when the surface is not spherical.\\
\noindent (3) The radius term $f^2(\sigma_0)\pi\sup l(S(0))/4$ is influenced by the behavior of the conformal factor $f$ within the entire foliation and thus can be interpreted as the 'global deviation' the system from being spherical.

The sum  \be\label{extraenergy}\frac{1}{8\pi}\int_\Sigma n^\sigma[(n_\sigma k)_{max}-n_\sigma k]\ud\Sigma+\frac{f^2(\sigma_0)\pi}{4}\sup l(S(0))\ee is the energy required to balance the non-sphericality when producing a trapped surface. By maximum principle, for $0\leq \sigma\leq \sigma_0$, $f(\sigma)\geq f(\sigma_0)$. Then the radius term $f^2(\sigma_0)\pi\sup l(S(0))/4 $ and the hoop term $f^{2}(\sigma_0)\int_0^{\sigma_0} C(s)\ud s$ are both less than the  terms in the sufficient condition for TS in terms of $M(\Sigma )$ \cite{M91},  $ f^2(0)\pi\sup l(S(0))/4:=\pi\mbox{rad}(0)/4$ and $\int_0^{\sigma_0} f^2(s)C(s)\ud s$, respectively. However, the 'boundary effect' energy compensating terms $\int_\Sigma n^\sigma[(n_\sigma k)_{max}-n_\sigma k]\ud\Sigma/8\pi$ are the same. We obtain a sufficient condition for TS's, employing the Brown-York mass, that is finer than that implied by Theorem \ref{restBY} (cf. Remark beneath Theorem \ref{restBY}).

In spherical symmetry, both of the two terms in Eq. (\ref{extraenergy}) vanish, and $C(s)\equiv1$ and hence $f^{2}(\sigma_0)\int_0^{\sigma_0} C(s)\ud s$ equals the areal radius of the surface. The inequality (\ref{suff}) is sharp and it reduces to the result in \cite{OTXM}.

\section{Conclusions}\label{S6}
We have shown, using conformally flat geometries and a suitable foliation, that the Brown-York mass is bounded from above by the total rest mass. The more non-spherical is a surface, the more Brown-York energy must be compacted within to make it trapped. We employ a number of geometric inequalities in Euclidean space, that yield several necessary and sufficient conditions  for ATS's and pointwise trapped surfaces. These results hold true for a large class of non-spherical geometries whose metrics are conformal (with convex layer surfaces) to the flat metric, and for adapted foliations.

\begin{acknowledgements}
N. Xie is partially supported by the National Science Foundation of China (grants 11171328, 11121101, 11421061) and CSC Program.
\end{acknowledgements}


\begin{thebibliography}{20}
\bibitem{BY}J.D. Brown and J.W. York, Phys. Rev. D  {\bf 47}, 1407 (1993).
\bibitem{OTXM}
N. \'{O} Murchadha, R.-S. Tung, N. Xie, and E. Malec, Phys. Rev. Lett. {\bf 104}, 041101 (2010).
\bibitem{Sei}H.J. Seifert, Gen. Relativ. Grav. {\bf 10}, 1065 (1979).
   \bibitem{Thorne}
K.S. Thorne in {\it Magic without Magic} ed. J. Klauder (Freeman, San Francisco), 231-258 (1972).
\bibitem{HW} J. Hartle and D. Wilkins, Phys. Rev. Lett. {\bf 31}, 60 (1973).
\bibitem{ADM} R. Arnowitt, S. Deser, and C.W. Misner, `The dynamics of General Relativity' in {\it Gravitation: an introduction to current research} , Ed. L. Witten, (Wiley, NY, 1962); Gen. Relativ. Grav. {\bf 40}, 1997 (2008).
 \bibitem{BMO88}P. Bizon, E. Malec, and N. \'{O } Murchadha, Phys. Rev. Lett. {\bf 61}, 1147 (1988).
\bibitem{BMO89}P. Bizon, E. Malec, and N. \'{O } Murchadha, Class. Quantum Grav. {\bf 6}, 961 (1989).
\bibitem{Ea}D. Eardley, J. Math. Phys. {\bf 36}, 3004 (1995).
 \bibitem{JO1}J. Guven and N. \'{O } Murchadha, Phys. Rev. D {\bf 56}, 7658 (1997).
\bibitem{JO2}J. Guven and N. \'{O } Murchadha, Phys. Rev. D {\bf 56}, 7666 (1997).

   \bibitem{M94}E. Malec, Phys. Rev. D {\bf 49}, 6475 (1994).

\bibitem{MO04}E. Malec and \'{O } Murchadha, Class. Quantum Grav. {\bf 21}, 5777 (2004).


\bibitem{Za92}T. Zannias, Phys. Rev. D {\bf 45}, 2998 (1992).
\bibitem{Za93}T. Zannias, Phys. Rev. D {\bf 47}, 1448 (1993).

    \bibitem{M91a} E. Malec, Acta Phys. Pol. B {\bf 22}, 829 (1991).

    \bibitem{Kh}M. Khuri, Phys. Rev. D {\bf 80}, 124025 (2009).
 \bibitem{BK}H. Bray and M. Khuri,  Asian J. Math. {\bf 15}, 557 (2011).


\bibitem{SY2} R. Schoen and S.-T. Yau, Commun. Math. Phys. {\bf 90}, 575 (1983).
  \bibitem{Ja}P.S. Jang, J. Math. Phys. {\bf 19}, 1152 (1978).
   \bibitem{SY} R. Schoen, and S.-T. Yau, Commun. Math. Phys. {\bf 79}, 231 (1981).

  \bibitem{KM} P. Koc and E. Malec, Acta Phys. Pol. B {\bf 23}, 123 (1992).


\bibitem{M91} E. Malec, Phys. Rev. Lett. {\bf 67}, 949 (1991).
 \bibitem{BOM}  R. Beig and N. \'{O } Murchadha, Phys. Rev. Lett. {\bf 66}, 2421 (1991).
 \bibitem{Sz}L. Szabados, Living Rev. Relativity {\bf 12}, 4 (2009).
 \bibitem{Ni}L. Nirenberg, Comm. Pure Appl. Math. {\bf 6}, 337 (1953).

 \bibitem{PS} G. P\'{o}lya and G. Szeg\"o,  {\it Isoperimetric inequalities in mathematical physics}. Annals of Mathematics Studies {\bf 27}, (Princeton University Press, Princeton, 1951).
     \bibitem{Kl}W. Klingenberg, {\it A course in differential geometry}, Translated from
the German by David Hoffman. Graduate Texts in Mathematics {\bf 51}, (Springer-Verlag, New York-Heidelberg, 1978).
  \bibitem{Schn}R. Schneider, {\it Convex Bodies: The Brunn-Minkowski Theory}, (Cambridge University Press, Cambridge, 2013).
  \bibitem{Fl}E. Flanagan, Phys. Rev. D {\bf 44}, 2409 (1991).
    \bibitem{Se}J.M.M. Senovilla, Europhys. Lett. {\bf 81}, 20004 (2008).
    \bibitem{JCAP} J.C.A. Paiva,Bull. Belg. Math. Soc. Simon Stevin {\bf 4}, 373 (1997).
    \bibitem{Birk} G.D. Birkhoff, Trans. Amer. Math. Soc. {\bf 18}, 199 (1917).
\bibitem{BM89}
  P. Bizon and E. Malec, Phys. Rev. D {\bf 40}, 2559 (1989).
\bibitem{BMOM1990}P. Bizon, E. Malec, and N. \'{O } Murchadha, Class. Quantum Grav. {\bf 7}, 1953 (1990).
\bibitem{Pe}R. Penrose, Phys. Rev. Lett. {\bf 14}, 57 (1965).
\bibitem{Malec22}E. Malec, Acta Phys. Pol. B {\bf 22}, 347 (1991).









\end{thebibliography}
\end{document}